\documentstyle[12pt,aps]{revtex}%
\newcommand{\bce}{\begin{center}}
\newcommand{\ece}{\end{center}}
\newcommand{\be}{\begin{equation}}
\newcommand{\ee}{\end{equation}}
\newcommand{\bea}{\vspace{0.25cm}\begin{eqnarray}}
\newcommand{\eea}{\end{eqnarray}}

\def\PLA{{Phys. Lett.}  A }
\def\PRL{{Phys. Rev. Lett.} }
\def\PRA{{Phys. Rev.} A }

\def\PRD{{Phys. Rev.} D }

\begin{document} \draft

\begin{center}
{\bf {\LARGE On the generation and identification of optical Schr\"odinger
cats }}
\end{center}

\vspace{ .25cm}
\begin{center}
{M.Genovese \footnote{ \small  genovese@ien.it, tel. 390113919234, fax
390113919259}, C.Novero} 
\\[0pt]

Istituto Elettrotecnico Nazionale Galileo Ferraris \\[0pt]
Str. delle Cacce 91 \\[0pt]
I-10135 Torino, Italy
\end{center}

\vspace{ 2.cm} {\large  Abstract }
\vskip 0.7cm
We discuss the possibility of generating and detecting, by a tomographic
reconstruction of the Wigner function, a macroscopic superposition of two
coherent states. The superposition state is created  using a conditioned
measurement on the polarisation of a probe photon entangled to a coherent
state. The entanglement is obtained using a Kerr cell inserted in a
Mach-Zender interferometer. Some hint about generation of GHZ states is
given as well.

\vspace{ 1.cm} {\large   }
\vskip 0.7cm

\vskip 3cm

Quantum Mechanics represents nowadays one of the pillars of modern physics:
so far a huge amount of theoretical predictions deriving by this theory
have been confirmed by very accurate experimental data. No doubts can be
raised on the validity of this theory.
Nevertheless, even at one century since its birth, many problems related to
the interpretation of this theory persist, in particular concerning the
concept of measurement in Quantum Mechanics and the related problem of the
transition from a microscopic probabilistic world to a macroscopic
deterministic world perfectly described by classical mechanics. The
disappearence of superposition effects for macroscopic bodies is known as
"decoherence".
Besides the huge theoretical interest for understanding this phenomenon, a
clear comprehension  of decoherence is also mandatory for constructing
quantum computers and thus it is of primary relevance in studies concerning
quantum information.

Many different models have been proposed for describing decoherence
\cite{books}, but none of them has reached a general acceptance.

However, it is now emerging the possibility of realising experiments
concerning the transition region between quantum and classical mechanics.
This is obtained studying mesoscopic states \cite{leg,har}, i.e. systems
composed by a number of particles for which one expects an intermediate
behaviour between quantum and classical mechanics.

In particular a huge interest is devoted to the generation of linear
superpositions of "macroscopically" distinguishable states, dubbed "
Schr\"odinger cats", which would permit a direct study of decoherence. 
Many different schemes have been proposed for realising superpositions of
many photons states using optical amplifiers \cite{OPA} or cavities
\cite{cav}. However, the practical realisation of these schemes is very
difficult due to the necessity of creating a state with an average number
of photons  sufficiently high to be considered a mesoscopic system without
that it decoheres too rapidly for being detected. Furthermore one has to
find a smart scheme for detecting the superposition. Up to now only the
beautiful experiment of Brune et al. \cite{har} has succeeded in obtaining
a superposition state involving an atom and few photons in a cavity.

It remains therefore a large interest in proposing and studying new
configurations for realising Schr\"odinger cats.

In a recent paper \cite{nos} we have proposed a scheme addressed to the
creation of a superposition of two coherent states using a Kerr cell.

In this letter we would like to investigate in details the real possibility
of creating and recognising such a state.
 
The idea is to use  the set-up of fig.1, where a "signal" photon enters the
Polarising-Beam-Splitter I (PBS I) from port 1. Let us suppose that this
photon is in a superposition of vertical (V) and horizontal (H)
polarisation, which will take different paths, for example let us assume
the vertical one will follow path 2 and the horizontal path 3. 

Furthermore, a probe laser crosses the Kerr cell on the  arm 3 acquiring a
phase or not according if the photon crosses or not the cell.

Thus the entangled state:
\be
\vert \Psi \rangle = { \vert H \rangle  \vert \nu '\rangle + \vert V
\rangle  \vert \nu \rangle \over \sqrt {2}}
\label{eq:Psi}
\ee
is generated, where the coherent state $\vert \nu '\rangle$ differs in
phase from $\vert \nu \rangle$.

The two signal photon paths are then recombined on a second beam splitter
(BSII) and a polarisation measurement (P1) is performed on this photon on
the base at $45 ^o$ respect to the horizontal-vertical base. This is the
conditional measurement producing the Schr\"odinger cat: if the signal
photon is found to have a  $45 ^o$ polarisation, the coherent state is
projected into the superposition
\be
\vert \psi_+ \rangle = {  \vert \nu '\rangle +   \vert \nu \rangle \over
\sqrt {2}}
\ee
on the other hand if the orthogonal polarisation is detected, the
projection will be in 
\be
\vert \psi_- \rangle = {  \vert \nu '\rangle -   \vert \nu \rangle \over
\sqrt {2}}
\ee
A superposition of two "many photons"  states is thus obtained.

The one photon signal state can be easily produced, for example,  using
parametric down conversion (PDC) in a non-linear crystal. In this case the
second photon of the down-converted pair can be used as trigger.
Furthermore, if the same pulsed laser is used both for pumping the crystal
and for the Kerr effect, one can easily obtain a good timing  for the
crossing  of the Kerr cell for the signal photon and the coherent state.

Incidentally, the use as input of a photon from PDC in a Kerr cell allows
also the creation of a three photons GHZ entangled state \cite{GHZ}, which
is one of the three elements necessary for realising an optical quantum
computer \cite{QC}, together with single qubit operations, which are easily
implemented, and teleportation. For what concerns this last  , a
description of a scheme performing this operation using a Kerr cell appears
in Ref. \cite{TVF}. Using the same polarisation dependence of the Kerr
interaction of Ref. \cite{TVF}, where there is no effect except when both
the photons interacting in the Kerr cell have vertical polarization ($\vert
V \rangle \vert V \rangle \rightarrow \vert V \rangle \vert V \rangle e^{i
\phi}$), a GHZ state can be generated by the interaction of an entangled
pair of photons with a third one.
Let us assume, for example, of having generated, by PDC, an entangled state
of the form \cite{nos2}:
\be
\vert \Phi ^+ \rangle = { \vert H \rangle \vert H \rangle + \vert V \rangle 
\vert V \rangle \over \sqrt {2}}
\label{eq:Psi2}
\ee
A simple algebraic calculation shows that  the interaction in the Kerr
medium of the second photon of the pair with a third photon polarised at
$45 ^o$ (denoted by $\vert 45 \rangle$, whilst its orthogonal state will be
$\vert 135 \rangle$), is, for a phase shift $\phi= \pi / 2$,  the GHZ state:
\be
\vert \Psi _{GHZ} \rangle = { \vert H \rangle \vert H \rangle \vert 45
\rangle + \vert V \rangle 
\vert V \rangle \vert 135 \rangle \over \sqrt {2}}.
\label{eq:GHZ}
\ee 
Analogous results are easily derived for the other three Bell states.

Recently a GHZ state of three photons has been generated experimentally
\cite{Jian}. However, albeit very interesting for testing local realism,
this scheme is not well suited for quantum computation, because of low
efficiency. Our scheme could thus represent an alternative for generating
GHZ states, which offers an interesting opportunity  for applications to
quantum computation.   

The interest for these schemes derives by the fact that, although
admittedly very difficult,  the Quantum Non Demolition (QND) \cite{QNDth}
detection of a single photon is at present possible \cite{SI,SM}. QND
measurements of { \it welcher Weg} (which path) have already been achieved 
using 100 meter long optical fibers (see Imoto et al. and Levenson et al.
\cite{QNDexp}). Of course, the implementation
of the present schemes using such devices would be, even though not impossible
in theory, almost impossible in practice. The
recent discovery of new materials with very high Kerr coupling, could
however allow an easier and more realistic implementation of this experiment.
Two candidates as Kerr cell with ultra-high susceptibility to be used for
this scheme are the Quantum Coherent Atomic Systems (QCAS)  \cite{QCAS,SI}
and the Bose-Einstein condensate of ultracold (at nanoKelvin temperatures)
atomic gas \cite{BEC}.
These are recent great technical improvements which could permit the
realisation of small Kerr cells, capable of large phase shift, even with a
low-intensity probe. In fact, both exhibit extremely high
Kerr couplings compared to more traditional materials. In particular, the
QCAS is rather a simple system to be realised (for a review see
\cite{Arimondo})  and thus represents an ideal candidate to this role.
Incidentally, one can notice that Kerr coupling can be further enhanced by
enclosing the medium in a cavity \cite{Agarwal}.

A main advantage of the present configuration  respect to the Schr\"odinger
cat generation using cavities is given by the fact that the coherent states
superposition travels in  air between the Kerr cell and the detection
apparatus. Decoherence appears when at least one photon is lost permitting
the identification of the phase of the state. In cavities this is related
to the damping factor of the cavity itself and decoherence becomes rather
rapid when the number of photons is increased (being proportional to the
average photon number). In the case under investigation the losses in air
are rather small and could permit a much longer time before decoherence
takes over.

In order to identify the macroscopic superposition we suggest to perform a
homodyne measurement of the laser beam after the Kerr cell. Then the Wigner
function can be reconstructed with tomographic techniques
\cite{tomo1,tomo2}, the identification of a negative region of the Wigner
function allows the identification of the cat (for a comprehensive study of
quantum tomography see
\cite{Leon}). 

More in details, one measures the probability distribution $p(x,\phi)$ for
the field quadrature $x=1/2(a e^{-i \phi} + a ^{\dagger} e^{i \phi})$ at
various phases $\phi$, then, using the Radon  transform, the Wigner
function is reconstructed.  Experimentally, one obtains a table of
$p(x,\phi)$ in terms of $x$ and $\phi$. In order not to have troubles with
delicate cancellations of different elements in the summations, it is
convenient to fit the $x$ dependence at different $\phi$ and then use this
fit in integrals.

In the following we will study the real possibility of reconstructing the
Wigner function and observing a negative region, when experimental errors
are kept into account. For this purpose, we will use the reconstruction
technique  described in \cite{tomo2}, appropriately deteriorating the
calculated distribution $p(x,\phi)$ in order to simulate experimental
errors. In practice, we introduce a random error, simulating the
experimental error on each fit of  $p(x,\phi)$ at fixed $\phi$ for each
Montecarlo run.

We will consider different superpositions systematically in order to give a
panorama of the different situations.

Probably, a good compromise between having a sufficiently high number of
photons and keeping at the same time a sufficiently small decoherence time
could be realised with a coherent state with an average number of photons
about 5-10.
This represents a "mesoscopic" system where intermediate properties between
"classical" and quantum systems can be expected. Furthermore, one can hope
to be able to study decoherence at work for such kind of systems.
  
Let us consider of having realised a superposition
\be 
\vert \Psi_{cat} \rangle = \vert \sqrt{5} e^{i \theta } \rangle +  \vert
\sqrt{5} e^{-i \theta } \rangle 
\ee

We begin with the case $\theta = \pi /2$.
The Wigner function $W[Re(\alpha), Im(\alpha)]$ (plotted in fig.2) has an
absolute minimum for $W[0.3346,0]=-3.16$. A tomographic reconstruction
based on ten phase points for $p(x, \phi)$ (this function is shown for the
sake of exemplification in fig.3) is sufficient for reproducing this value
with an uncertainty less than 3\%. We have then considered the situation
where the distribution  $p(x, \phi)$ is deteriorated simulating an
experimental error around $\pm 25 \%$, the result of an average on ten
points calculated by a Montecarlo simulation is $ W[0.3346,0]=-3.08 \pm
0.29$: then even with such large errors bars we are able to reconstruct the
minimum negative value with only a $10 \%$ uncertainty. For the sake of
completeness, we have also repeated the evaluation with a simulated error
of 50 \%, the result is $ W[0.3346,0]=-2.84 \pm 0.84$. Even in this case
the negative part of the Wigner function would be recognisable (with a 30
\% relative error).  

Then we have investigated the case $\theta = 63 ^o$.
The Wigner function $W[Re(\alpha), Im(\alpha)]$ (plotted in fig.4) has an
absolute minimum for $W[0.8954,0]=-3.916$. Also in this case a tomographic
reconstruction based on ten phase points for $p(x, \phi)$ is sufficient for
reproducing this value with an uncertainty less than 3\%, albeit a larger
calculation time is needed. Again, we have considered the situation where
the distribution  $p(x, \phi)$ is deteriorated simulating an experimental
error around $\pm 25 \%$, the result of the Montecarlo simulation
(averaging 10 points) is $ W[0.8954,0]=- 3.83 \pm 0.48$: we are thus able
to reconstruct the minimum negative value with only a $ 12.5 \%$ uncertainty.
For the sake of completeness, we have also investigated the tomographic
reconstruction of the local minimum $W[0.157,0]=-0.890$. An average of ten
Montecarlo simulated points gives $W[0.157,0]=-0.910 \pm 0.193$, with a
relative error of $21.3 \%$.

When the relative phase between the two coherent states is further reduced
the tomographic reconstruction of the Wigner function requires a very large
calculation time; however, up to a phase difference of 0.1 radiants the
negative region of the Wigner function can still be  identified (e.g. for
$\theta = 0.2$ radiants the minimum is $W[2.687,0]=-0.679$).

Finally,  we have also investigated the case where the average number of
photons is 10: a part the need of a longer calculation time, the same
situation as the previous case is found. 

In summary, the results of our numerical simulation show that the
tomographic reconstruction of the Wigner function is such that even with
$25 \%$, or larger, errors on the homodyne reconstruction of $p(x, \phi)$
the cat can be easily identified.
We think thus that this technique can be an interesting tool for
recognising an optical Schr\"odinger cat realised as superposition of two
coherent state with different phase.

In conclusion, we have shown that an optical Schr\"odinger cat
(superposition of two coherent states) can be generated using a Mach-Zender
interferometer 
crossed by a one photon state, where on one of the paths of the
interferometer a Kerr cell is inserted. A laser crosses the cell as well
and the coherent states superposition is generated by a conditional
measurement on the one photon state.
The produced state is robust against decoherence.

We have also discussed the experimental possibility of recognising the cat,
once it has been created, using a tomographic technique for reconstructing
its Wigner function. Our numerical simulations show that this
reconstruction is such that the cat can be recognised even if rather large
experimental errors affect the homodyne detection. 

Altogether, our results show that the proposed  scheme  for generating and
detecting an optical superposition of macroscopic states can be efficient,
also thanks to the appearance of new materials with very high Kerr
couplings, as Bose condensate. We think that this scheme deserves attention
for an experimental realisation of it. 
   
\vskip 1cm
{\bf Acknowledgements}

\noindent We would like to acknowledge support of ASI under contract LONO
500172, of  MURST via special programs "giovani ricercatori" Dip. Fisica
Teorica Univ. Torino and of Istituto Nazionale di Fisica Nucleare.


\vskip 1cm

{ \noindent {\bf References}}


\begin{enumerate}

\bibitem{books} see D. Bohm and B.J. Hiley, "The undivided universe", 1994
Routledge; R. Omn\'es, "The intepretation of quantum mechanics", 1990
Princeton; P. Pearle quant-ph 9901077; R. Bonifacio, Nuov. Cim. B 114
(1999) 473 
and references therein.

\bibitem{leg} A.J. Legett, Phys. Rev. B 30 (1984) 1208.

\bibitem{har} M. Brune et al., \PRL  77 (1996) 4887; L. Davidovich et al.,
\PRA 53 (1996) 1295.
  
\bibitem{OPA} F. De Martini et al., \PRA 60 (1999) 1636.

\bibitem{cav} see for example Yurke and Stoler, \PRL 57 (1986) 13; S. Song
et al., \PRA 41 (1990) 5261; M. Brune et al., \PRA 45 (1992) 5193; P.
Tombesi and D. Vitali, \PRL 77 (1996) 411; P. Goetsch, P. Tombesi and D.
Vitali, \PRA 54 (1996) 4519; D. Vitali, P. Tombesi and P. Grangier, Appl.
Phys. B 64 (1997) 249.

\bibitem{nos} M. Genovese and C. Novero, Phys. Rev. A 61 032102 (2000).
  
\bibitem{SM} B.C. Sanders and G.J. Milburn, \PRA 39 (1989) 694.

\bibitem{SI} Q.A. Turchette et al., \PRL 75 (1995) 4710; H. Schmidt and A.
Imamoglu, Opt. Lett. 21 (1996) 1936; A. Imamoglu et al., \PRL 79 (1997) 1467.


\bibitem{GHZ} D.M. Greenberger et al., \PRA 82 (1999) 1345 and ref.s therein.

\bibitem{QC} D. Gottesman and I.I. Chuang, quant-ph 9908010.

\bibitem{TVF} D. Vitali et al., quant-ph 0003082.

\bibitem{Jian} Jian-Wei Pan, D. Bouwmeester, M. Daniell, H. Weinfurter and
A. Zeilinger, Nature 403 (2000) 515.

\bibitem{nos2} G. Brida, M. Genovese, C. Novero and E. Predazzi, \PLA 268
(2000) 12 and ref.s therein.

\bibitem{QNDth} V.B. Braginsky and Y.I. Vorontsov, Sov. Phys. Uzp. 17
(1975) 644;
W.G. Unruh, \PRD 19 (1979) 2888; V.B. Braginsky et al., Science 209 (1980)
547; N. Imoto et al., \PRA 32 (1985) 2287. 

\bibitem{QNDexp} M.D. Levenson et al., \PRL 57 (1986) 2473; N. Imoto et al.,
Opt. Comm. 61 (1987) 159; A. La Porta et al., \PRL 62 (1989) 28; P.
Grangier et al., \PRL 66 (1991) 1418. For a review see: V. B. Braginsky and
F.Y. Khalili,
Rev. Mod. Phys. 68 (1996) 1.

\bibitem{Kim} Y.-H. Kim, S.P. Kulik, Y.H. Shih and M.O. Scully,
quant-ph/9903047.


\bibitem{QCAS} U. Rathe et al., \PRA 47 (1993) 4994; M.M. Kash et al., \PRL
82 (1999) 5229. 

\bibitem{BEC} L. Vestergaard Hau et al., Nature 397 (1999) 594. S.E. Harris
and L.V. Hau, \PRL 82 (1999) 4611.

\bibitem{Arimondo} E. Arimondo, in Progress in optics XXXV, E. Wolf
editor, Elsevier Science 1996, pag. 257.


\bibitem{Agarwal} G.S. Agarwal, Opt. Comm. 72 (1989) 253.

\bibitem{tomo1} K. Vogel and H. Risken, \PRA 40 (1989) 2847;
D.T. Smithey et al., \PRL 70 (1993) 1244; G. M. D'Ariano, U. Leonhardt and
H. Paul, \PRA  52 (1995) R1801.

\bibitem{tomo2} G.M. D'Ariano, C. Macchiavello and G.A. Paris, \PRA 50
(1994) 4298, \PLA 195 (1994) 31 and Nuov. Cim. 110B (1995) 237.

\bibitem{Leon} U. Leonhardt, "Measuring the quantum state of light",
Cambridge University press, Cambridge, 1997.

\bibitem{Chi3} see e.g. R. L. Sutherland, "Handbook of nonlinear optics",
ed. M. Dekker 1996.


\vfill \eject

{\bf Figure captions }

\vskip 0.5cm 

fig.1) Scheme of the proposed experiment. The signal photon enters gate 1
of the polarising beam splitter PBS. A Kerr cell (K) is on one arm of the
interferometer. A probe laser crosses the cell. The signal beam is measured
by the photo-detectors D1 and D2 at the out gates of the interferometer.
Finally a homodyne detection is performed on the laser after exiting the
Kerr cell. 


fig.2) Contour plot of the  Wigner function $W[Re(\alpha), Im(\alpha)]$  of
the state $
\vert \sqrt{5} e^{i \pi/2 } \rangle +  \vert \sqrt{5} e^{-i \pi/2 }  \rangle$.

fig.3) Three dimensional plot of $p(x,\phi)$ for the state $
\vert \sqrt{5} e^{i \pi/2 } \rangle +  \vert \sqrt{5} e^{-i \pi/2 }
\rangle$. The x-axis is the quadrature variable $x$, on the y-axis appear
the points $n$, where $n$ varies between 0 and 10 and is related to $\phi$
through $\phi= n * (\pi/2) / 10$.

fig.4) Contour plot of the  Wigner function $W[Re(\alpha), Im(\alpha)]$  of
the state $
\vert \sqrt{5} e^{i 63^o } \rangle +  \vert \sqrt{5} e^{-i 63^o } \rangle$.
   


\end{enumerate}
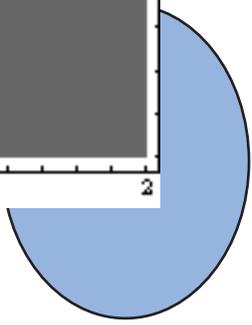
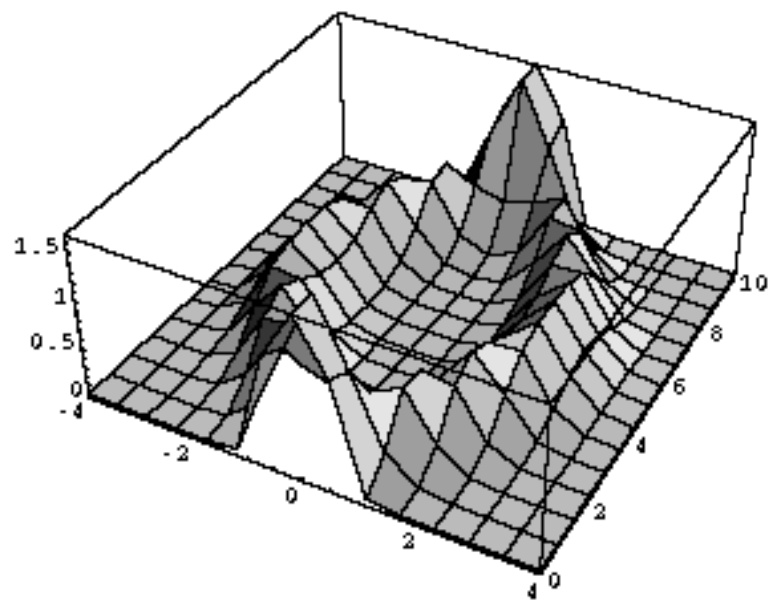
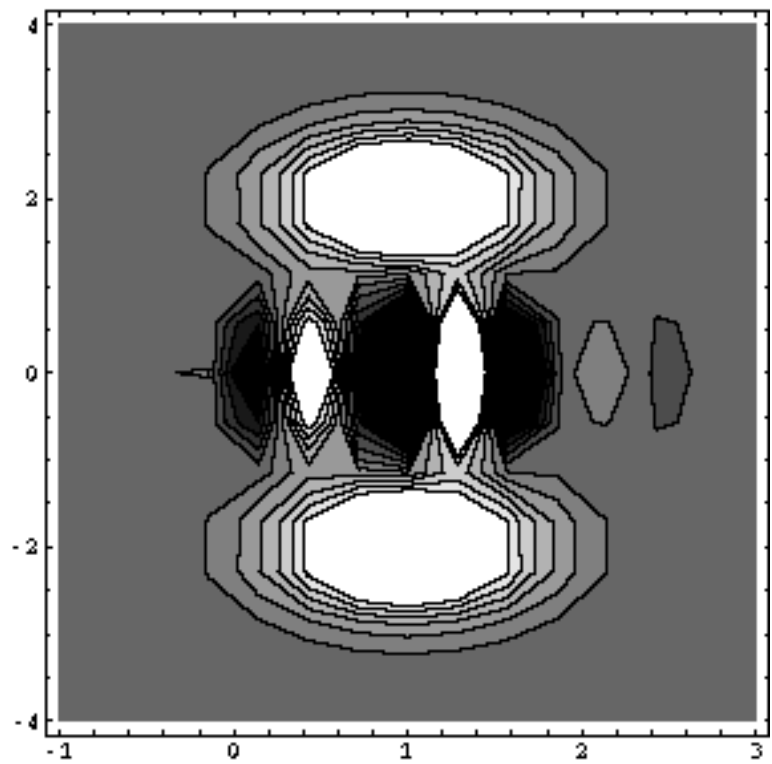
\end{document}